\documentclass[sigconf, 9pt]{acmart}
\usepackage{pifont}
\usepackage{multirow}
\usepackage{subfigure}
\usepackage{graphicx}
\usepackage{balance}
\usepackage[linesnumbered,ruled,slide]{algorithm2e} % for algorithm

\SetCommentSty{mycommfont}

\AtBeginDocument{%
  }

\setcopyright{acmlicensed}
\copyrightyear{2018}
\acmYear{2018}
\acmDOI{XXXXXXX.XXXXXXX}
\acmConference[Conference acronym 'XX]{Make sure to enter the correct
  conference title from your rights confirmation email}{June 03--05,
  2018}{Woodstock, NY}
\acmISBN{978-1-4503-XXXX-X/2018/06}

\begin{document}
\newcommand{\name}{{TiInsight}\xspace}
\newcommand{\tisql}{{TiSQL}\xspace}
\newcommand{\tichart}{{TiChart}\xspace}
\newcommand{\papername}{{\name: A SQL-based Automated Exploratory Data Analysis System through Large Language Models}\xspace}
\title{\papername}

\settopmatter{authorsperrow=5}

\author{Jun-Peng Zhu}
\affiliation{
  \institution{Northwest A\&F University, PingCAP}
  \country{}
}

\author{Boyan Niu}
\affiliation{
  \institution{PingCAP, China}
  \country{}
}

\author{Peng Cai}
\affiliation{
  \institution{East China Normal University, China}
  \country{}
}

\author{Zheming Ni}
\affiliation{
  \institution{PingCAP, China}
  \country{}
}

% \author{Jianwei Wan}
% \affiliation{
%   \institution{PingCAP, China}
%   \country{}
% }

\author{Kai Xu}
\affiliation{
  \institution{PingCAP, China}
  \country{}
}

\author{Jiajun Huang}
\affiliation{
  \institution{PingCAP, China}
  \country{}
}

\author{Shengbo Ma}
\affiliation{
  \institution{PingCAP, China}
  \country{}
}

\author{Bing Wang}
\affiliation{
  \institution{PingCAP, China}
  \country{}
}

\author{Xuan Zhou}
\affiliation{
  \institution{East China Normal University, China}
  \country{}
}

\author{Guanglei Bao}
\affiliation{
  \institution{PingCAP, China}
  \country{}
}

\author{Donghui Zhang}
\affiliation{
  \institution{PingCAP, China}
  \country{}
}

\author{Liu Tang}
\affiliation{
  \institution{PingCAP, China}
  \country{}
}

\author{Qi Liu}
\affiliation{
  \institution{PingCAP, China}
  \country{}
}

\renewcommand{\shortauthors}{Jun-Peng Zhu et al.}

\begin{abstract}
The SQL-based exploratory data analysis has garnered significant attention within the data analysis community.
The emergence of large language models (LLMs) has facilitated the paradigm shift from manual to automated data exploration.
However, existing methods generally lack the ability for cross-domain analysis, and the exploration of LLMs capabilities remains insufficient.
This paper presents \name, a SQL-based automated cross-domain exploratory data analysis system.
First, \name offers a user-friendly GUI enabling users to explore data using natural language queries.
Second, \name offers a robust cross-domain exploratory data analysis pipeline: hierarchical data context  (i.e., HDC) generation, question clarification and decomposition, text-to-SQL (i.e., \tisql), and data visualization (i.e., \tichart).
Third, we have implemented and deployed \name in the production environment of PingCAP and demonstrated its capabilities using representative datasets.
\end{abstract}





\maketitle

\section{Introduction}

The exploratory data analysis~\cite{chat2query} with SQL has garnered significant attention within the data analysis community.
This process involves crafting SQL queries to achieve specific data exploration goals and visually presenting the results using dedicated tools.
Users iteratively perform these operations until satisfactory data exploration results are achieved.
This process involves two key pain points that impact the efficiency and quality of data exploration: (1) proficiency in SQL and (2) recommending appropriate data visualization charts.
Several text-to-SQL and table-to-chart approaches have been proposed to tackle these pain points.
The rise of large language models (LLMs) provides new opportunities to address these challenges.
As far as we know, state-of-the-art (SOTA) approaches have their limitations:

\textbf{(1) Limited cross-domain generalization in text-to-SQL.}
The LLM-based text-to-SQL methods have demonstrated superiority over traditional approaches.
However, these methods~\cite{zhang2024finsql} frequently require fine-tuning LLMs to adapt to varying data domains.
This approach requires well-labeled data and is time-consuming.
In practice, fine-tuning is not effective for data exploration across real-world data domains.
Alternative LLM-based approaches employ diverse prompt techniques to enhance text-to-SQL accuracy.
However, these techniques are predominantly benchmark-oriented and lack sufficient accuracy~\cite{zhang2024finsql} for real-world scenarios.

\textbf{(2) Inability to effectively address unclear user intentions in text-to-SQL.}
In real-world EDA scenarios, users often struggle to articulate their exploration goals when working with unfamiliar datasets.
This leads to the failure of SOTA text-to-SQL methods to generate correct SQL statements, and in many cases, no SQL statements are generated.
For instance, in the PingCAP production environment, users frequently inquire, ``What is the growth rate?''.
This query may lack a time parameter; a more precise formulation would be, ``What is the growth rate for the current year?''.
In the worst-case scenario, users may struggle to clearly articulate their intent at the beginning of a data analysis task.
Accurately addressing these questions requires not only the semantic parsing capabilities of LLMs but also robust data analysis skills to infer intent beyond the explicit query.
Existing SOTA text-to-SQL approaches struggle to generate SQL in this context, let alone address accuracy issues.

\textbf{(3) Overly complex table-to-chart recommendation process in table-to-chart.}
Existing table-to-chart methods~\cite{zhou2021table2charts} typically rely on reinforcement learning or deep neural networks, which tend to be overly complex.
Some methods~\cite{lee2021lux} necessitate users mastering a specific visualization query language, significantly restricting their applicability in end-to-end EDA systems.
Numerous user studies found that most charts follow specific visualization rules.
For example, the pie chart is suitable for displaying the proportion or distribution of categorical data.
In particular, data analysts have accumulated numerous heuristic rules for data visualizations, but these lack a straightforward way of applying them to visualization recommendations.
The advent of LLMs presents new opportunities for rule-based data visualization.

To address these limitations, we propose \name, a SQL-based end-to-end automated cross-domain exploratory data analysis system.
\name enables end-to-end data exploration without necessitating prior expertise from users.
Driven by user queries in natural language, \name seamlessly transitions from natural language input to visualization exploration results.
This system introduces the following unique and innovative features.
To begin with, \name provides an intuitive GUI that allows users to interact with exploratory datasets through natural language input seamlessly.
Second, \name incorporates HDC generation capability, facilitating efficient execution of cross-domain data exploration tasks.
The third key feature, built upon HDC, is that \name incorporates the efficient text-to-SQL tool \tisql.
Additionally, \name incorporates a rule-based data visualization tool \tichart.
With these capabilities at its disposal, \name showcases its value across a wide range of cross-domain data analysis scenarios.

Through this demo, SIGMOD attendees can experience TiInsight in action. After the attendee specifies a dataset, they can (1) use the HDC component to leverage the LLM's context summarization capability to characterize the user's dataset, (2) propose the question they want to obtain from the dataset using natural language. TiInsight then uses the question clarification and decomposition component to refine and decompose the user's question. During this process, the visitors can get the clarified question through the GUI, (3) automatically generate the corresponding SQL statement utilizing TiSQL, and execute it against the database for each sub-question, and finally (4) generate visual charts corresponding to the results.
In this paper, we demonstrate \name in two representative real-world scenarios.
The first is Financial, a dataset within the target industry business at PingCAP during the Proof of Concept (PoC) phase.
This dataset provides detailed information on the U.S. Federal Reserve's federal funds rate, as well as data on key economic indicators.
The second is Bird~\cite{bird}. % which is an innovative dataset that investigates the influence of large database contents on text-to-SQL parsing.
The system is available at \textcolor{blue}{\url{https://www.tiinsight.chat}}. The demo video is available at \textcolor{blue}{\url{https://youtu.be/JzYFyYd-emI}}.
\begin{figure}[!t]
\centering
\includegraphics[width=0.35\textwidth]{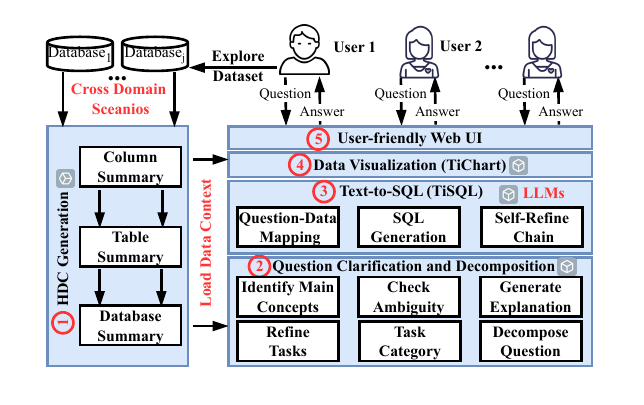}
\vspace{-0.3cm}
\caption{Overall architecture of \name.}
\vspace{-0.7cm}
\label{fig:architecture}
\end{figure}

\vspace{-0.5cm}
\section{System Overview} \label{sec:architecture}
\name offers a natural language interface enabling end users to conduct online exploratory data analysis.
Figure~\ref{fig:architecture} illustrates the architecture of \name.
The architecture of \name can be roughly divided into the following five components:

\noindent\textbf{\underline{\ding{172} HDC Generation}}.
We design the HDC generation component, which uses LLMs to efficiently generate the hierarchical data context of the database schema online at the beginning of exploration.
The LLMs are powerful for summarizing text content from massive datasets.
It can capture the most significant content from the database schema to guide the subsequent exploratory process.
This component allows us to understand the data provided by the user across various hierarchies of the database schema.

\noindent\textbf{\underline{\ding{173} Question Clarification and Decomposition}}.
We decode and clarify questions from different users by leveraging LLMs.
Specifically, this component resolves ambiguous user intentions and provides possible explanations.
Then, TiInsight determines whether the user's task should be decomposed into a series of sub-tasks.

\noindent\textbf{\underline{\ding{174} Text-to-SQL (\tisql)}}.
We employ \tisql to convert user questions into SQL statements.
First, \tisql needs to map the user question to specific tables and columns while providing the values for SQL conditional statements.
In this process, we propose a two-stage mapping method from coarse-grained to fine-grained.
Subsequently, \tisql employs LLMs to generate SQL statements.
We carefully design the prompt of \tisql.
However, the SQL statements generated in this step still contain errors.
As a result, \tisql employs a self-refinement chain to enhance the generated SQL.

\noindent\textbf{\underline{\ding{175} Data Visualization (\tichart)}}.
In \name, we utilize \tichart to visualize the user's query results to the fullest extent.
A key challenge in the data visualization process arises from the complexity of user tasks, which can generate numerous sub-tasks, leading to uncertainty in the final query (e.g., attributes and data).
This greatly amplifies the complexity of visualizing the resulting data.
To address this challenge, \tichart combines rule-based methods with LLMs to recommend suitable visualization types.

\noindent\textbf{\underline{\ding{176} User-friendly Web UI}}.
The user interface of \name offers a wide range of capabilities.
Users have the flexibility to import their data for exploration purposes.
Additionally, it features an intuitive interface that enables users to interact with their data through natural language.
\name includes a \textit{bookmark} feature that facilitates users in comparing different data exploration results.
Additionally, \name supports switching between different LLMs via the web UI.

We briefly introduce the two components of \name in the following sections. More technical details and experimental results are provided in our research paper~\cite{zhu2024towards}.

\begin{figure}[!t]
\centering
\includegraphics[width=0.25\textwidth]{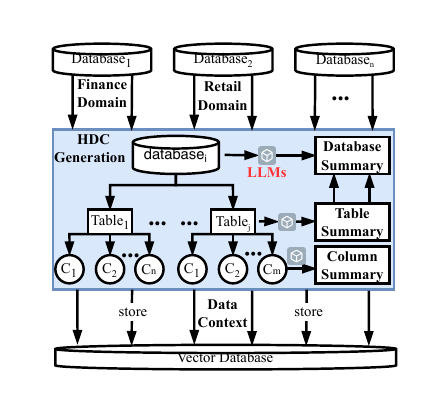}
\vspace{-0.3cm}
\caption{Overview of hierarchical data context.}
\vspace{-0.7cm}
\label{fig:hdc}
\end{figure}

% \vspace{-0.4cm}
\section{Hierarchical Data Context Generation} \label{sec:hdc}

The hierarchical data context is designed primarily to enhance the understanding of the database schema, which includes column summaries, table descriptions, table-to-table relationships, and an overall database summary.
This design is derived from experience during the PoC stage at PingCAP.
In PingCAP's data analysis business, data analysts first collect and organize relevant database schema information when encountering unfamiliar datasets to gain a better understanding of the database.
We leveraged this experience and fully automated the process using LLMs, as illustrated in Figure~\ref{fig:hdc}.

\noindent\underline{\textbf{Column Summary}}. A column-by-column summarization approach would result in high latency due to repeated calls to the LLMs.
Additionally, using the LLM for each column individually is inefficient, leading to significant token wastage and increased costs.
To address these challenges, we propose a parallel, grouping-based approach to mitigate this issue by vertically partitioning the table.
First, HDC organizes columns into fixed sets through vertical partitioning.
This limitation is primarily due to the context window constraints of these LLMs.
Each group is subsequently allocated threads from a thread pool before the LLM initiates the column description process.
Comments may be included in the schema information provided by the schema for each column.
After the aforementioned process, we append the column comments to each column description.
Additionally, for each column, we randomly select several rows (i.e., three rows in this paper) to include in the column description, aiding the LLM in understanding the column's value type and other relevant information.
Finally, the LLM receives the column content and dynamically constructs the CoT prompt to facilitate the column summary.

\noindent\underline{\textbf{Table Description}}. To create a comprehensive table description, we need to gather additional information, including the primary key, key attributes, table type, table entity, and natural language description.
Specifically, the LLM may identify multiple primary keys, and we prompt it to select the most likely one.
This primary key should better reflect the importance and uniqueness of the table.
Key attributes play a vital role in understanding the purpose and content of the table.
We prompt the LLM to identify and list a maximum of five key attributes from the table presented in this paper.
In addition, we categorize the table type as either dimension, bridge, or fact.
This classification is especially crucial for the OLAP analysis.
Finally, we identify the main entity that the table focuses on.

\noindent\underline{\textbf{Table Relationship}}. We propose a two-stage approach for identifying the table relationships.
To determine the referential integrity of a specific table with other tables, in \textbf{stage~1 (i.e., coarse-grained search)}, we retrieve \textit{similar\_count} tables from the vector database using the specified table description.
In \textbf{stage~2 (i.e., fine-grained exploration)}, we input the specified table along with the table retrieved in stage~1 into the LLMs for fine-grained identification by CoT prompt.
We iterate over each table to establish an integrity relationship for all tables within the database. Given $n$ tables, the time complexity to explore all tables for each specified table is $O(n^2)$. Using a coarse-grained search reduces this complexity, allowing us to achieve a time complexity of $O(n)$.

\noindent\underline{\textbf{Database Summary}}. To facilitate efficient data exploration across multiple databases, we propose extracting representative \textbf{entities} for each database, enabling a \textbf{database summary} based on these entities.
A database consists of numerous tables with complex relationships.
Following the influence maximization principle, a table with a higher number of relationships to other tables is considered more important within the database.
Therefore, we propose an entity extraction method that leverages the number of relationships between tables.
The approach selects the top N tables with the most relationships from the current database.
A prompt is then constructed using these tables and their summaries, and the LLM is employed to infer their entities.
Finally, the inferred entities are saved in the vector database.
We infer the database summary from the entities using the LLM with a CoT prompt.

\vspace{-0.3cm}
\section{\tisql: Text-to-SQL based on HDC} \label{sec:tisql}

We propose a schema filtering framework based on the map-reduce paradigm to filter tables and columns.
\tisql begins with a coarse-grained search, retrieving the top $N$ relevant tables from the vector database based on the clarified question and cosine similarity.
Then, \tisql performs fine-grained filtering to exclude irrelevant tables and columns.
To prevent the associated table and column summaries from exceeding the LLM's context limit, \tisql divides these tables into more granular groups of related tables.
Subsequently, map-reduce is employed to process the corresponding block utilizing idle threads.
In the map phase, the prompt is dynamically constructed using the table and column summaries of the current group and the clarified task.
The LLM, combined with the CoT prompt, is then used to identify the relevant tables and required columns.
In the reduce phase, we apply the reduce mode for each fine-grained table group to merge the retrieved tables and columns.
\tisql utilizes both the HDC and the map-reduce framework to streamline the LLM prompt, significantly reducing the complexity of schema linking.

The SQL statements generated from the above procedure may still contain errors; therefore, we propose a self-refinement chain.
After generating SQL statements from the LLM, the results are initially fed into an explain-refine process, where potential errors can be identified.
The \textit{EXPLAIN} statement is a widely used tool in databases that provides feedback on potential issues in SQL statements without requiring their execution.
After refining the SQL statements using the \textit{EXPLAIN} statement, certain errors may only become apparent during execution; therefore, we integrate the \textit{EXECUTION} statement into the \tisql refinement chain.
Execute the SQL statements and relay any feedback from error information to the LLM for correction.
The final SQL output is generated after iterating the self-refinement chain.
% \vspace{-0.3cm}
\section{Demonstration Scenarios}  \label{sec:demo}

We introduce a web application to the participants and demonstrate \name through two real-world scenarios: the Financial dataset and the Bird dataset.

\vspace{-0.3cm}

\subsection{Scenario 1: GUI Interface}

\begin{figure}[!h]
\centering
\includegraphics[width=0.48\textwidth]{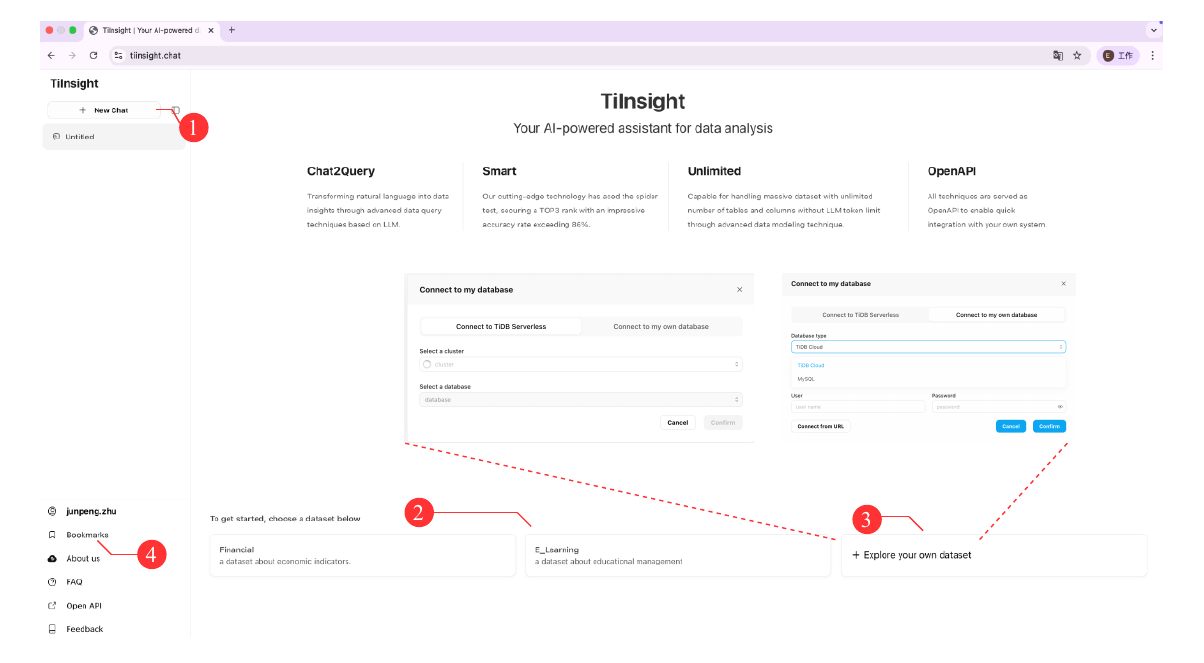}
\vspace{-0.5cm}
\caption{An illustration of a GUI interface.}
\vspace{-0.5cm}
\label{fig:demo1}
\end{figure}

To improve usability for users, \name provides a conversational GUI to facilitate user-system interaction. When the user enters ``https://www.tiinsight.chat/'' in the browser, the page shown in Figure~\ref{fig:demo1} is displayed. The interface includes the \textit{New Chat} button (\ding{182}), two default TiInsight datasets (\ding{183}), a dataset exploration interface (\ding{184}), and a left-side toolbar (\ding{185}) containing bookmarks, API documentation, and additional features. In the following demonstration, we illustrate the system's functionality using the two default datasets provided by TiInsight. First, a new data exploration and analysis session is created by clicking the \textit{New Chat} button and selecting the \textit{Financial} dataset.

\subsection{Scenario 2: Financial Dataset}

\begin{figure}[!h]
\centering
\includegraphics[width=0.48\textwidth]{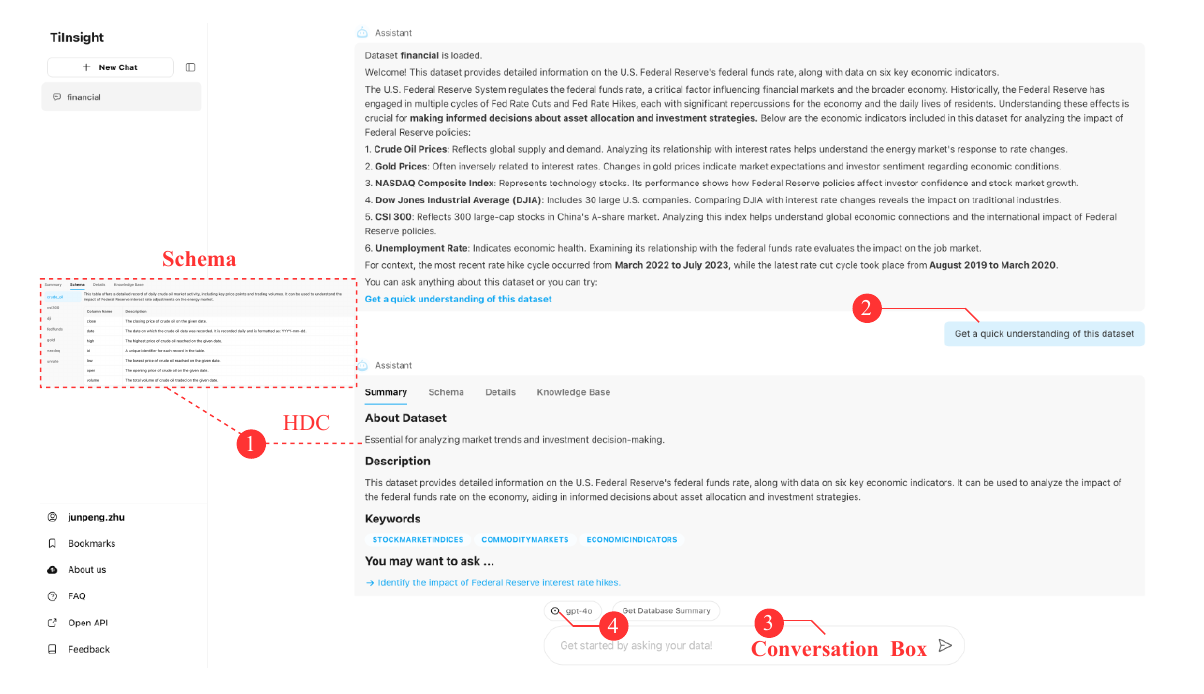}
\vspace{-0.5cm}
\caption{An illustration of an EDA task on a financial dataset.}
\vspace{-0.5cm}
\label{fig:demo2}
\end{figure}

 We use the natural language question ``Identify the impact of Federal Reserve interest rate hikes.'' and an incomplete question ``Changes in oil prices?'' as demonstration examples. Figure~\ref{fig:demo2} illustrates the key features of the user interface.

\noindent\textbf{Step 1: HDC Generation}. First, \name provides the HDC (\ding{182}) of the user's dataset, including a summary, a description, keywords, table information, and more. Users can obtain HDC by the NL question ``Get a quick understanding of this dataset'' (\ding{183}).

\noindent\textbf{Step 2: EDA Task/Goal}. Users can either input their question in the  \textit{Conversation Box} (\ding{184}) or begin with a data exploration question suggested by \name. The \textit{Conversation Box} feature enables users to transition between various LLMs seamlessly (\ding{185}). Next, \name leverages the HDC of the current dataset to interpret the user's question and refine the exploration task.

\begin{figure}[!h]
    \centering
    \includegraphics[width=0.48\textwidth]{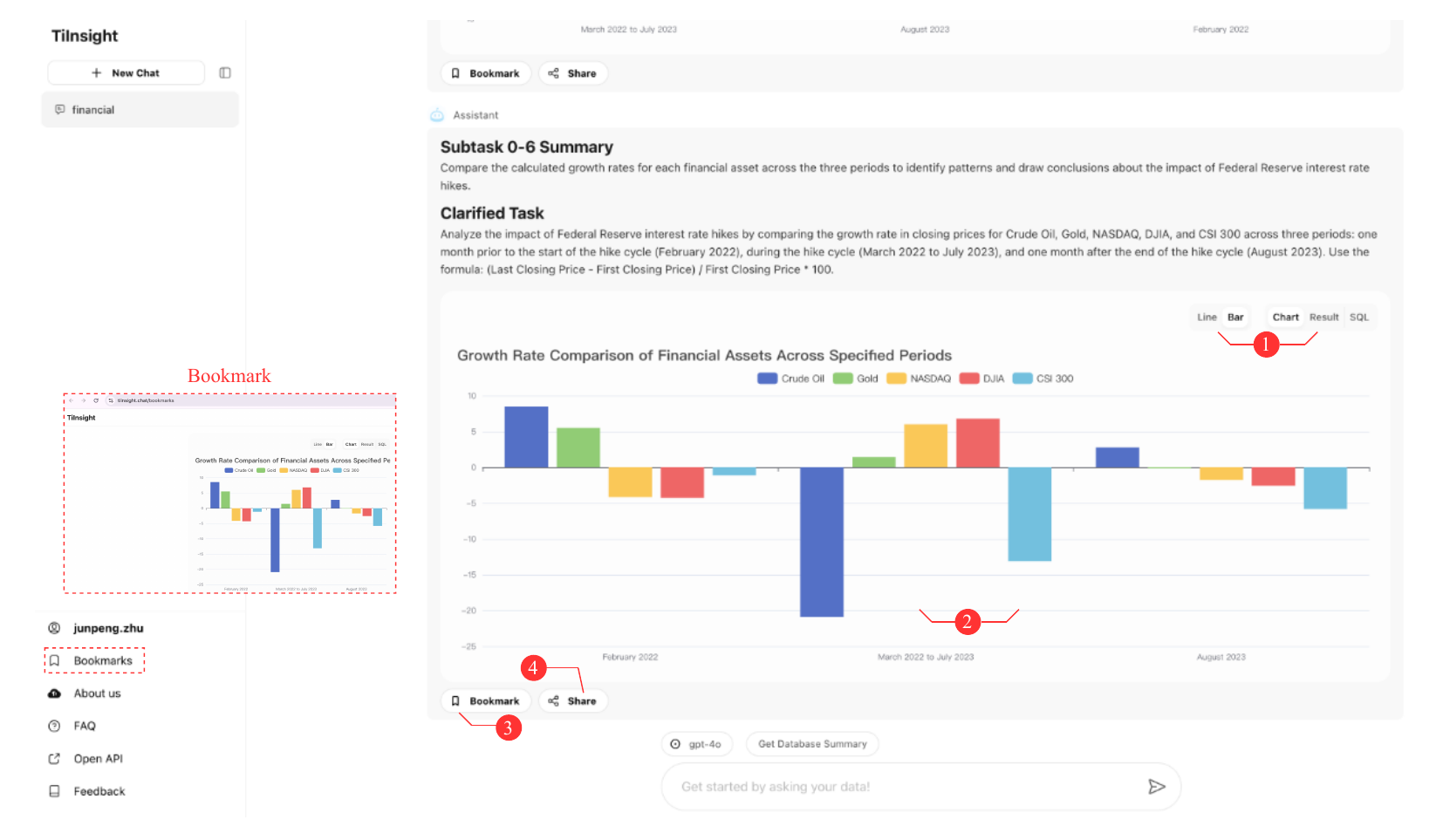}
    \vspace{-0.3cm}
    \caption{An illustration of an EDA result on a financial dataset.}
    \vspace{-0.5cm}
    \label{fig:demo3}
    \end{figure}

\noindent\textbf{Step 3: EDA Results}. Based on the exploration task refined by \name, the \tisql tool generates the corresponding SQL statements while \tichart selects the most suitable chart type to present the analysis results by visualization as shown in Figure~\ref{fig:demo3}-\ding{182}.
\tichart may recommend multiple visualization types for the results, allowing users to switch between them as needed (\ding{183}).
For each exploration result, \name provides a ``Bookmark'' button, enabling users to save the result to bookmarks for easy comparison (\ding{184}).
Users can click the \textit{Bookmarks} button in the left-side toolbar to compare different EDA results.
Additionally, users can click the \textit{share} button (\ding{185}) to disseminate their analysis results.

\vspace{-0.3cm}
\subsection{Scenario 3: Bird Dataset}

In this demonstration, we have opted to utilize the exploration task ``List each test result and its count in descending order of count.'' as an exemplar to explore the capabilities of \name.
For a more comprehensive understanding of features, please refer to our video for more details.

\section{Conclusion}  \label{sec:conclusion}

In this paper, we introduced \name, a SQL-based automated cross-domain exploratory data analysis system.
We presented the implementation of \name and demonstrated its novel features on a user-friendly interface, powerful hierarchical data context (i.e., HDC) generation, \tisql, \tichart, and exploration efficiency.

\balance
\bibliographystyle{ACM-Reference-Format}
\bibliography{ref}

@INPROCEEDINGS{chat2query,
  author={Zhu, Jun-Peng and Cai, Peng and Niu, Boyan and Ni, Zheming and Xu, Kai and others},
  booktitle={ICDE},
  title={Chat2Query: A Zero-Shot Automatic Exploratory Data Analysis System with Large Language Models},
  year={2024},
  volume={},
  number={},
  pages={5429-5432},
  }

@article{zhang2024finsql,
  title={FinSQL: Model-Agnostic LLMs-based Text-to-SQL Framework for Financial Analysis},
  author={Zhang, Chao and Mao, Yuren and Fan, Yijiang and Mi, Yu and Gao, Yunjun and Chen, Lu and Lou, Dongfang and Lin, Jinshu},
  journal={arXiv preprint arXiv:2401.10506},
  year={2024}
}

@inproceedings{zhou2021table2charts,
  title={Table2Charts: recommending charts by learning shared table representations},
  author={Zhou, Mengyu and Li, Qingtao and He, Xinyi and Li, Yuejiang and Liu, Yibo and Ji, Wei and Han, Shi and Chen, Yining and Jiang, Daxin and Zhang, Dongmei},
  booktitle={SIGKDD},
  pages={2389--2399},
  year={2021}
}

@article{lee2021lux,
  title={Lux: always-on visualization recommendations for exploratory dataframe workflows},
  author={Lee, Doris Jung-Lin and Tang, Dixin and others},
  journal={VLDB},
  volume={15},
  number={3},
  pages={727--738},
  year={2021},
  publisher={VLDB Endowment}
}

@article{bird,
  title={Can llm already serve as a database interface? a big bench for large-scale database grounded text-to-sqls},
  author={Li, Jinyang and Hui, Binyuan and Qu, Ge and Yang, Jiaxi and Li, Binhua and Li, Bowen and Wang, Bailin and Qin, Bowen and Geng, Ruiying and Huo, Nan and others},
  journal={NIPS},
  volume={36},
  year={2024}
}

@article{zhu2024towards,
author = {Zhu, Jun-Peng and Niu, Boyan and Cai, Peng and Ni, Zheming and Wan, Jianwei and Xu, Kai and others},
title = {Towards Automated Cross-Domain Exploratory Data Analysis through Large Language Models},
year = {2025},
issue_date = {August 2025},
publisher = {VLDB Endowment},
volume = {18},
number = {12},
issn = {2150-8097},
url = {https://doi.org/10.14778/3750601.3750629},
journal = {Proc. VLDB Endow.},
month = aug,
pages = {5086–5099},
numpages = {14}
}

\end{document}